\pdfoutput=1
\documentclass[%
aps,
 amsmath,amssymb,
 reprint,%
]{revtex4-1}

\usepackage[pdftex]{graphicx}
\usepackage{dcolumn}% Align table columns on decimal point
\usepackage{bm}% bold math
\usepackage{amsmath}

\begin{document}

\title{The effect of intrinsic point defects on ferroelectric polarization behavior of SrTiO$_3$ }% Force line breaks with \\

\author{Konstantin Klyukin}
\email {kklyukin2@unl.edu}
\affiliation{Department of Chemical and Biomolecular Engineering, University of Nebraska-Lincoln, Lincoln, NE 68588, United States}

\author{Vitaly Alexandrov}
\email {valexandrov2@unl.edu}
\affiliation{Department of Chemical and Biomolecular Engineering, University of Nebraska-Lincoln}
\affiliation{Nebraska Center for Materials and Nanoscience, University of Nebraska-Lincoln, Lincoln, NE 68588, United States}%

\date{\today}% It is always \today, today,
             %  but any date may be explicitly specified

\begin{abstract}
The effect of a variety of intrinsic defects and defect clusters in bulk and thin films of SrTiO$_3$ on ferroelectric polarization and switching mechanism is investigated by means of density-functional-theory (DFT) based calculations and the Berry phase approach. Our results show that both the titanium Ti$_\mathrm{Sr}^{\bullet \bullet}$ and strontium  Sr$_\mathrm{Ti}^{''}$ antisite defects induce ferroelectric polarization in SrTiO$_3$, with the Ti$_\mathrm{Sr}^{\bullet \bullet}$ defect causing a more pronounced spontaneous polarization and higher activation barriers of polarization reversal than Sr$_\mathrm{Ti}^{''}$. The presence of oxygen vacancies bound to the antisite defects can either enhance or diminish polarization depending on the configuration of the defect pair, but it always leads to larger activation barriers of polarization switching as compared to the antisite defects with no oxygen vacancies. We also show that the magnitude of spontaneous polarization in SrTiO$_3$ can be tuned by controlling the degree of Sr/Ti nonstroichiometry. Other intrinsic point defects such as Frenkel defect pairs and electron small polarons also contribute to the emergence of ferroelectric polarization in SrTiO$_{3}$.

\end{abstract}

 \maketitle

%\tableofcontents
\section{INTRODUCTION}

Switchable polarization in ferroelectric materials due to the orientation of dipoles by an external electric field is central to various energy and information storage technologies including sensors and actuators \cite{MRS:8739827},  electro-optic devices \cite{wessels2007ferroelectric,xiong2014active,george2015lanthanide}, ferroelectric field-effect transistors for non-volatile memories \cite{scott2007applications,garcia2014ferroelectric}. In the past years it has been revealed that ferroelectric polarization is not exclusive to polar-materials and can be induced throughout the non-ferroelectric layer of the heterostructure by combining a non-ferroelectric oxide such as SrTiO$_3$ with a ferroelectric oxide, e.g., BaTiO$_{3}$,\cite{Rabe2003} or even with another non-ferroelectric oxide, e.g., LaCrO$_{3}$. \cite{ADMI:ADMI201500779} Moreover, the emergence of net ferroelectric polarization was recently demonstrated for nanometer-thick films of SrTiO$_{3}$ \cite{lee2015emergence} where this effect was attributed to electrically induced alignment of polar nanoregions that can naturally form because of the presence of intrinsic defects in SrTiO$_{3}$ crystals. It was previously demonstrated that intrinsic defects such as the antisite Ti defects can form in the bulk phase of Ti-rich SrTiO$_{3}$, generate local polarization around the antisite Ti center due to an off-center displacement and might contribute to the appearance of polar nanoregions \cite{lee2015emergence,choi2009role} in a manner similar to extrinsic defects.\cite{Bianchi1995}

Native point defects in perovskite-structured SrTiO$_3$ were studied extensively in the past both experimentally and theoretically with the largest emphasis being placed on the oxygen vacancy as the most prominent point defect in SrTiO$_3$ that affects a wide range of material properties including electronic and optical behavior.\cite{PhysRevB.75.121404,Alexandrov2009,Yamada2009,Demkov2012,choi2013anti,Janotti2014,PhysRevB.47.8917} SrTiO$_3$ point defect chemistry, thermodynamics and kinetics of defect formation and diffusion were also investigated in great detail. \cite{muller2004atomic,Merkle2008,kotomin2011confinement, PhysRevLett.105.226102,liu2014composition} For example, oxygen vacancies serve as a source of $n$-type conductivity that can vary with oxygen partial pressure and are responsible for insulator-to-metal transition\cite{PhysRevB.47.8917}. Oxygen vacancies are also known to play a key role in the resistive switching process under applied electric field due to their low activation energies of diffusion.\cite{PhysRevB.75.121404,ADMA:ADMA200602915,oxygen_switching,switching} Also, it is well established that point defects including oxygen vacancies play a critical role in mediating polarization switching in ferroelectrics by controlling the local polarization stability, acting as pinning sites for domain-wall motion and ultimately defining the mechanism and kinetics of polarization switching.\cite{chu2005elastic,kalinin2010defect}

The impact of intrinsic point defects including oxygen vacancies on polarization switching phenomenon in SrTiO$_3$ is much less understood. In this study we carry out a systematic investigation of the effect of native defects in bulk and thin-film SrTiO$_3$ on ferroelectric polarization and polarization reversal at a single defect level by means of first-principles electronic structure calculations.

\section{COMPUTATIONAL METHOD}

First-principles calculations are performed within the density functional theory (DFT) formalism using the projector augmented wave (PAW) potentials \cite{kresse1999ultrasoft} as implemented in the Vienna Ab initio Simulation Package (VASP).\cite{vasp} The PAW potentials for Sr, Ti, O and Ru contain 10, 12, 6 and 14 valence electrons,  respectively, that is, Sr: $4s^2$$4p^6$$5s^2$, Ti: $3s^23p^64s^23d^2$ O: $2s^22p^4$  and Ru: $4p^65s^14d^7$. The generalized gradient approximation  Perdew-Burke-Ernzerhof (GGA-PBE) exchange-correlation functional\cite{GGA-PBE} is employed in the modified form for solids PBEsol\cite{PBEsol} along with a plane wave cutoff energy of 400 eV. The rotationally invariant PBEsol$+U$ approach is adopted with $U_{eff}$ = 4.36 eV on the Ti 3$d$ orbitals. The ions are relaxed by applying a conjugate-gradient algorithm until the Hellmann-Feynman forces are less than 20 meV/{\AA} with an optimized lattice constant of 3.903 \AA. The $3\times 3\times 3$ Monkhorst-Pack $k$-mesh is used for the Brillouin zone integration for a 3$\times$3$\times$3 supercell, while the mesh was adjusted for other supecells to provide a similar \emph{k}-point density in each direction.

\begin{figure}[h]
  \centering
  \includegraphics[width=0.4\textwidth]{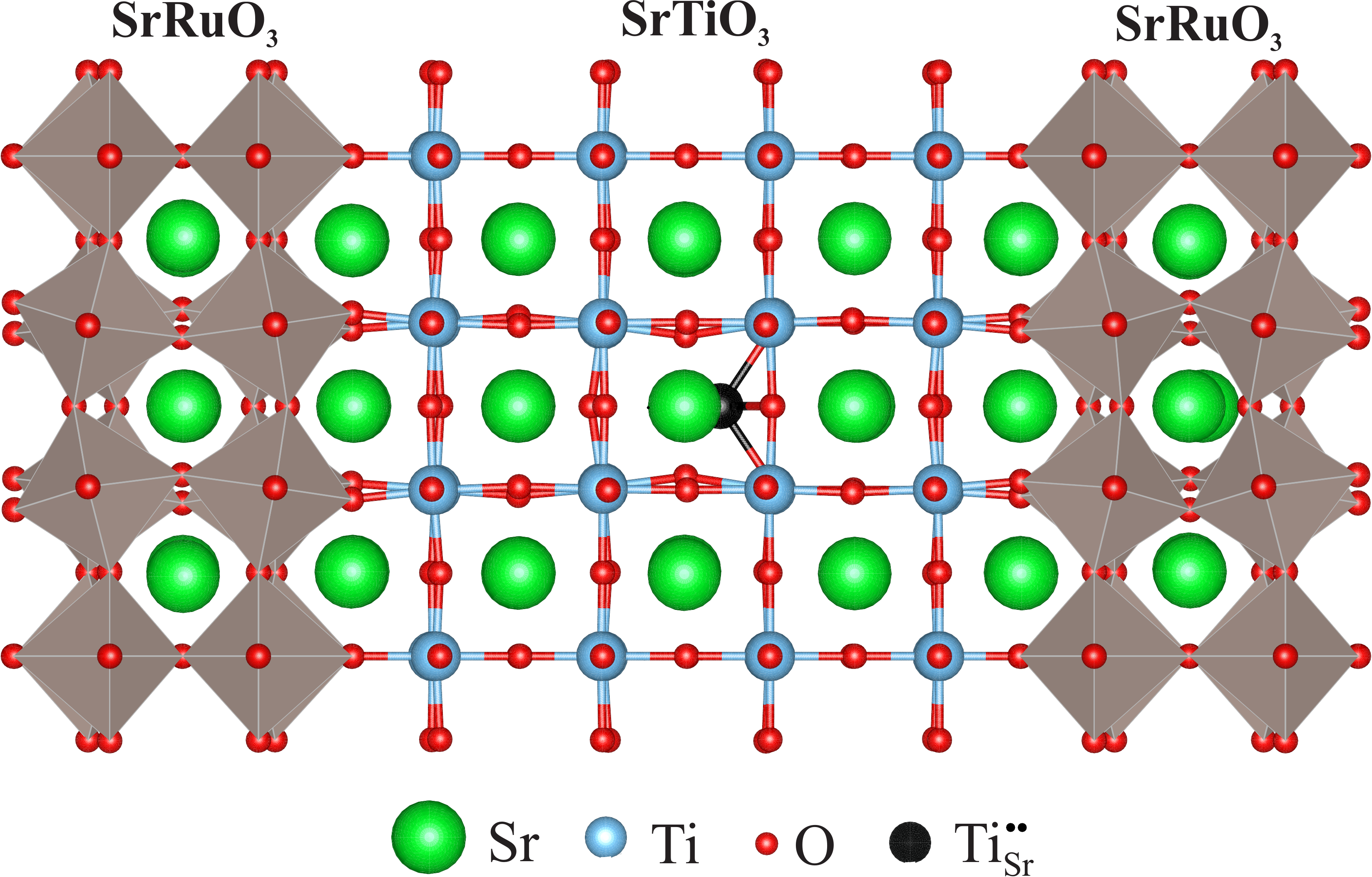}
  \caption{The atomic structure of SrTiO$_3$/SrRuO$_3$ thin films with the antisite Ti$_\mathrm{Sr}^{\bullet \bullet}$ defect in the middle of the supercell which induces polarization along the [100] direction.}
  \label{film}
\end{figure}
To investigate the influence of intrinsic defects and defect clusters on the polarization properties of SrTiO$_3$, we construct a $3\times3\times3$ supercell consisting of 135 atoms for the bulk calculations and a 3$\times$3$\times$7 multilayered structure  comprised of four SrTiO$_3$ and three SrRuO$_3$ layers for the thin-film calculations (see Figure \ref{film}).  The Berry-phase approach\cite{berry} within the modern theory of polarization is employed to calculate polarization properties. According to this approach the spontaneous polarization is defined as the difference in polarization between the polar and non-polar (centrosymmetric) reference states.\cite{spaldin2012beginner} To estimate the polarization switching barriers we calculate the migration energy profile E$_m$ along the minimum energy path between two polarization states ($P_-$ and $P_+$) using the climbing image nudged elastic band (CI-NEB) method.\cite{NEB} To denote the SrTiO$_3$ point defects we adopt the Kr\"{o}ger-Vink nomenclature.\cite{Kroeger1964,Merkle2008}

\section{RESULTS AND DISCUSSION}

\subsection{Ti$_\mathrm{Sr}^{\bullet \bullet}$ antisite defect}

We start by considering the titanium-strontium Ti$_\mathrm{Sr}^{\bullet \bullet}$ antisite defect where the Ti$^{4+}$ ion occupies a site on the Sr$^{2+}$ sublattice. This defect was predicted to be the dominant defect in SrTiO$_{3}$ along with the oxygen vacancy V$_\mathrm{O}$ under Ti-rich conditions.\cite{choi2009role,liu2014composition} To find the most stable atomic configuration for Ti$_\mathrm{Sr}^{\bullet \bullet}$ we examine the atomic structures with the Ti atom shifted along the [100], [110] and [111] crystallographic directions. A large Ti$_\mathrm{Sr}^{\bullet \bullet}$ off-centering of 0.78 {\AA} along the [100] direction is found to be the most energetically favorable with an energy gain of 0.48 eV with respect to the non-shifted configuration, in agreement with previous estimates.\cite{lee2015emergence,choi2009role} The displaced Ti atom forms four Ti$-$O bonds of length 2.20 {\AA}, whereas in defect-free SrTiO$_3$ the Ti$-$O bond distances are 1.95 {\AA}.  The atomic configuration with the shifted Ti$_\mathrm{Sr}^{\bullet \bullet}$ can be considered as an electric dipole comprised of a negatively charged Sr vacancy and a positively charged Ti interstitial which induces the electric polarization. Using the Berry phase method we estimate the average polarization of the supercell $P^{100}$(Ti$_\mathrm{Sr}^{\bullet \bullet}$) to be 16.8 $\mu$C/cm$^2$. In full agreement with previous calculations,\cite{lee2015emergence} we find that despite the large off-centering of the Ti$_\mathrm{Sr}^{\bullet \bullet}$, its local dipole moment is relatively small due to a small Born effective charge of 1.72 (see Table \ref{tab1}). Consequently, the overall dipole moment is dominated by the induced dipole moments in the surrounding cells rather than by the dipole moment of the antisite Ti atom which accounts for about 8.1\% of the total dipole moment of the supercell.

\begin{figure}[!h]
  \centering
  \includegraphics[width=0.4\textwidth]{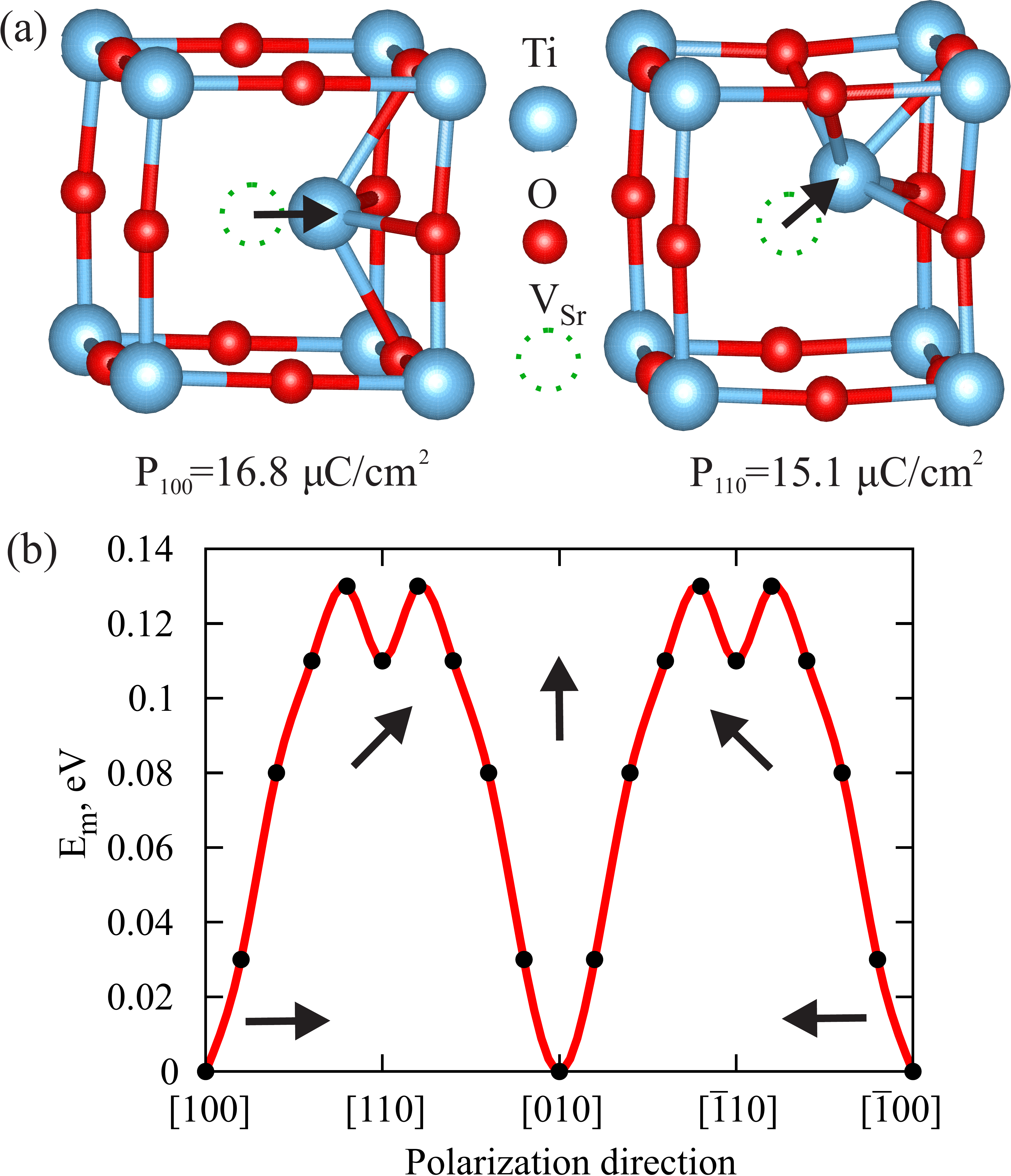}
  \caption{(a) Atomic structures of SrTiO$_3$ with the antisite Ti$_\mathrm{Sr}^{\bullet \bullet}$ defect for two polarization states with Ti$_\mathrm{Sr}^{\bullet \bullet}$ shifted along the [100] and [110] directions. (b) Migration energy profile between polarization states caused by the Ti$_\mathrm{Sr}^{\bullet \bullet}$ defect. Polarization reversal from [100] to [$\bar{1}$00] is achieved via the metastable polarization states with the [110] and [$\bar{1}$01] directions.}\label{antiTi}
\end{figure}

We also estimate the migration energy barriers for [100] $\to$ [$\bar{1}$00] polarization switching and find that the barrier for the direct switching between the two polarization states is rather large (0.48 eV), while the two-step migration via the intermediate state [110] is characterized by the barrier of only 0.13~eV (see Figure \ref{antiTi}). For this metastable state the average supercell polarization $P^{110}$(Ti$_\mathrm{Sr}^{\bullet \bullet}$) = 15.1 $\mu$C/cm$^2$.

The influence of oxygen vacancies on SrTiO$_3$ polarization properties is not well understood at the \textit{ab initio} level despite the predominant role of this defect in SrTiO$_3$ defect chemistry. Previous theoretical studies suggested that Ti$_\mathrm{Sr}^{\bullet \bullet}$ and V$_\mathrm{O}$ together with V$_\mathrm{Sr}^{''}$ should be the most thermodynamically stable defects in SrTiO$_{3}$ under Ti-rich conditions,\cite{choi2009role,liu2014composition,Janotti2014} while Ti-rich environment is predicted to be energetically more favorable that excess SrO in SrTiO$_{3}$.\cite{liu2014composition} Calculated formation energies as a function of Fermi level indicate that the doubly charged V$_\mathrm{O}^{\bullet \bullet}$ should be more stable than the singly charged V$_\mathrm{O}^{\bullet}$ and neutral V$_\mathrm{O}^{\times}$ even in $n$-type SrTiO$_{3}$ in which the Fermi level is close to the bottom of the conduction band.\cite{choi2009role,PhysRevB.90.085202} It is expected that the presence of the positively charged oxygen vacancies in the vicinity of the Ti$_\mathrm{Sr}^{\bullet \bullet}$ defect may change the dipole moment induced by Ti$_\mathrm{Sr}^{\bullet \bullet}$.
\begin{figure}[!h]
  \centering
    \includegraphics[width=0.4\textwidth]{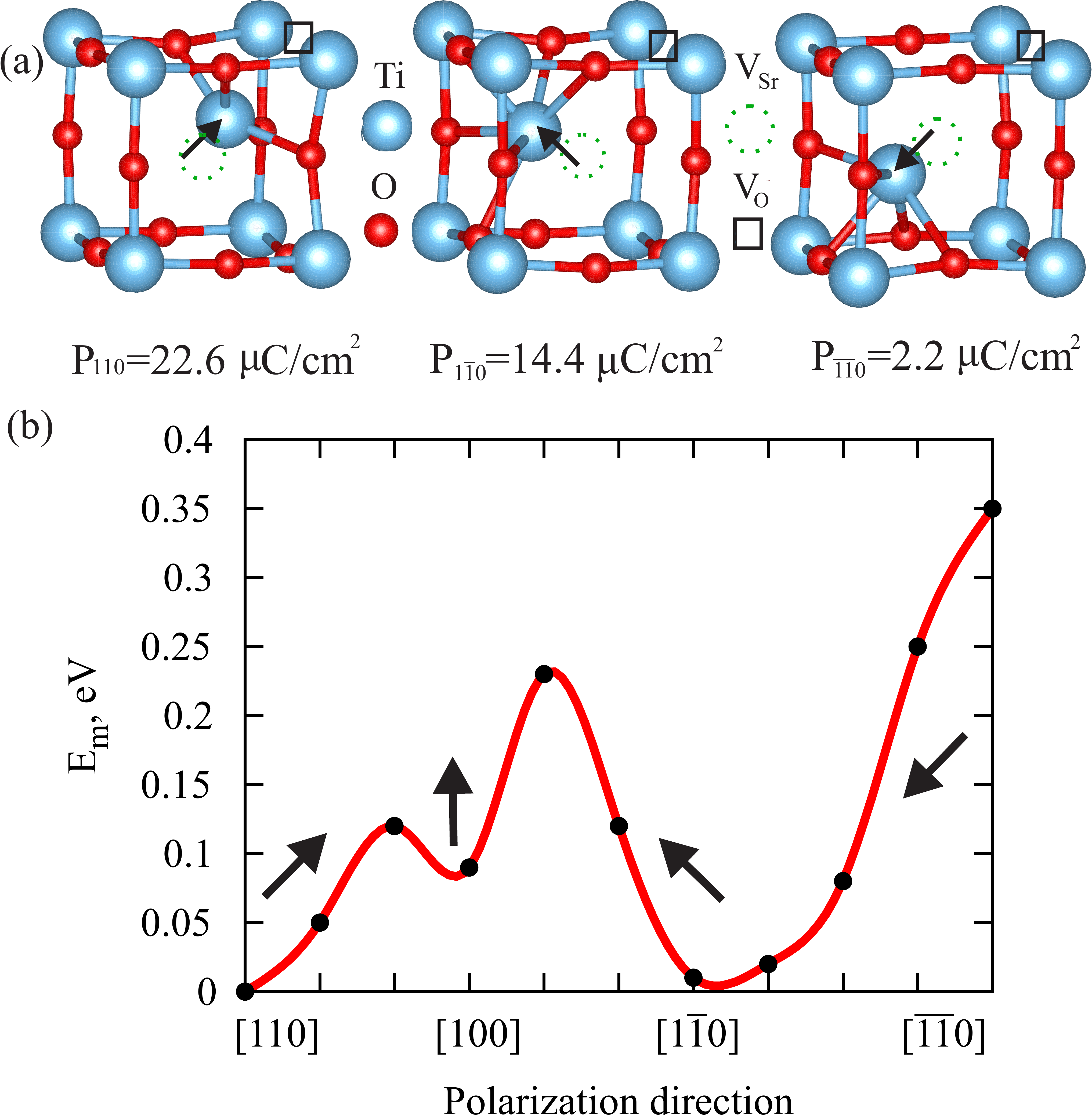}
   \caption{(a) Atomic structures of SrTiO$_3$ with Ti$_\mathrm{Sr}^{\bullet \bullet}$ and V$_\mathrm{O}^{\bullet \bullet}$ for polarization states with Ti$_\mathrm{Sr}^{\bullet \bullet}$ shifted along the [110], [1$\bar{1}$0] and [$\bar{1}$$\bar{1}$0] directions. (b) Migration energy profiles between polarization states caused by the Ti$_\mathrm{Sr}^{\bullet \bullet}$ and V$_\mathrm{O}^{\bullet \bullet}$ defects. Polarization switching from [110] to [1$\bar{1}$0] can be achieved via the metastable polarization state with the [100] direction.}
   \label{antiTi_charge}
\end{figure}

First, our calculations reveal a negative binding energy of about -0.4 eV between V$_\mathrm{O}^{\bullet \bullet}$ and Ti$_\mathrm{Sr}^{\bullet \bullet}$ indicating that the formation of the defect complex is energetically favored over the isolated defects. To examine different atomic arrangements between these defects, we displace Ti$_\mathrm{Sr}^{\bullet \bullet}$ with respect to V$_\mathrm{O}^{\bullet \bullet}$ as shown in Figure \ref{antiTi_charge}. We find that the most stable configuration is non-magnetic and characterized by a Ti$_\mathrm{Sr}^{\bullet \bullet}$ off-centering of 0.79 {\AA} along the [110] direction towards the vacancy exhibiting enhanced polarization $P^{110}$(Ti$_\mathrm{Sr}^{\bullet \bullet}$$-$V$_\mathrm{O}^{\bullet \bullet}$) = 22.6 $\mu$C/cm$^2$ as compared to the Ti$_\mathrm{Sr}^{\bullet \bullet}$ case with no oxygen vacancy. We also find that a slightly less favorable (by 0.02 eV) spin-polarized configuration with a magnetic moment of 2~$\mu_{B}$ has a much lower polarization $P^{110}$(Ti$_\mathrm{Sr}^{\bullet \bullet}$$-$V$_\mathrm{O}^{\bullet \bullet}$) = 5.61 $\mu$C/cm$^2$ caused by a much less pronounced off-centering of 0.43 {\AA}.

The non-symmetrical state $P_-$  is characterized by a reduced polarization $P^{1\bar{1}0}$(Ti$_\mathrm{Sr}^{\bullet \bullet}$$-$V$_\mathrm{O}^{\bullet \bullet}$) = 14.4 $\mu$C/cm$^2$ caused by a 0.81 {\AA} off-centering. Such a decrease relative to the most stable $P^{110}$ state could be explained by the opposite directions of dipoles formed by V$_\mathrm{Sr}^{''}$-Ti$_\mathrm{Sr}^{\bullet \bullet}$ and V$_\mathrm{Sr}^{''}$-V$_\mathrm{O}^{\bullet \bullet}$. The switching barrier between these two polarization states is computed to be 0.24 eV, which is twice higher than for  Ti$_\mathrm{Sr}^{\bullet \bullet}$ with no oxygen vacancy. A displacement along the [$\bar{1}\bar{1}0$] direction leads to a substantially diminished polarization $P^{\bar{1}\bar{1}0}$(Ti$_\mathrm{Sr}^{\bullet \bullet}$$-$V$_\mathrm{O}^{\bullet \bullet}$) = 2.2 $\mu$C/cm$^2$ and a greater switching barrier.

\begin{figure}[!h]
  \centering
  \includegraphics[width=0.4\textwidth]{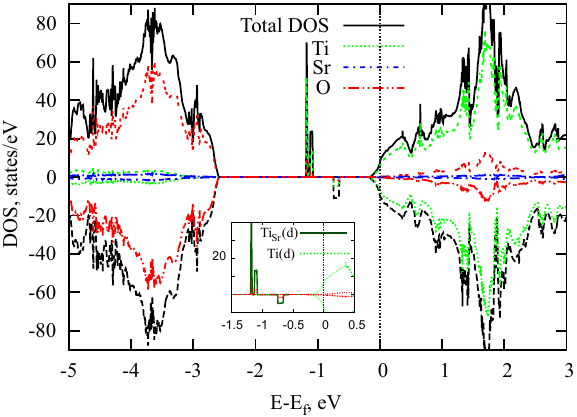}
  \caption{Density of electronic states calculated for the Ti$_\mathrm{Sr}^{\bullet \bullet}$$-$V$_\mathrm{O}^{\times}$ defect complex. The Fermi level corresponds to zero.}\label{dos}
\end{figure}

We next analyze the Ti$_\mathrm{Sr}^{\bullet \bullet}$$-$V$_\mathrm{O}^{\times}$ defect complex since neutral V$_\mathrm{O}^{\times}$ may have the formation energy only slightly higher than those of the positive charge states in the $n$-type region.\cite{Janotti2014} We find that the complex is stable with an estimated binding energy of about -0.35 eV, but is characterized by the metallic behavior and no polarization can be given. In this case one electron of the antisite defect moves to the conduction band forming a metallic state near the Fermi level while the second electron forms a localized in-gap state (Figure \ref{dos}).  In relation to polarization properties this suggests that the formation of the Ti$_\mathrm{Sr}^{\bullet \bullet}$$-$V$_\mathrm{O}^{\times}$ defect complexes may contribute to the interplay between polarization and resistive switching behaviors in Ti-rich SrTiO$_{3}$.

\subsection{Sr$_\mathrm{Ti}^{''}$ antisite defect}

\begin{figure}[!h]
  \centering
  \includegraphics[width=0.4\textwidth]{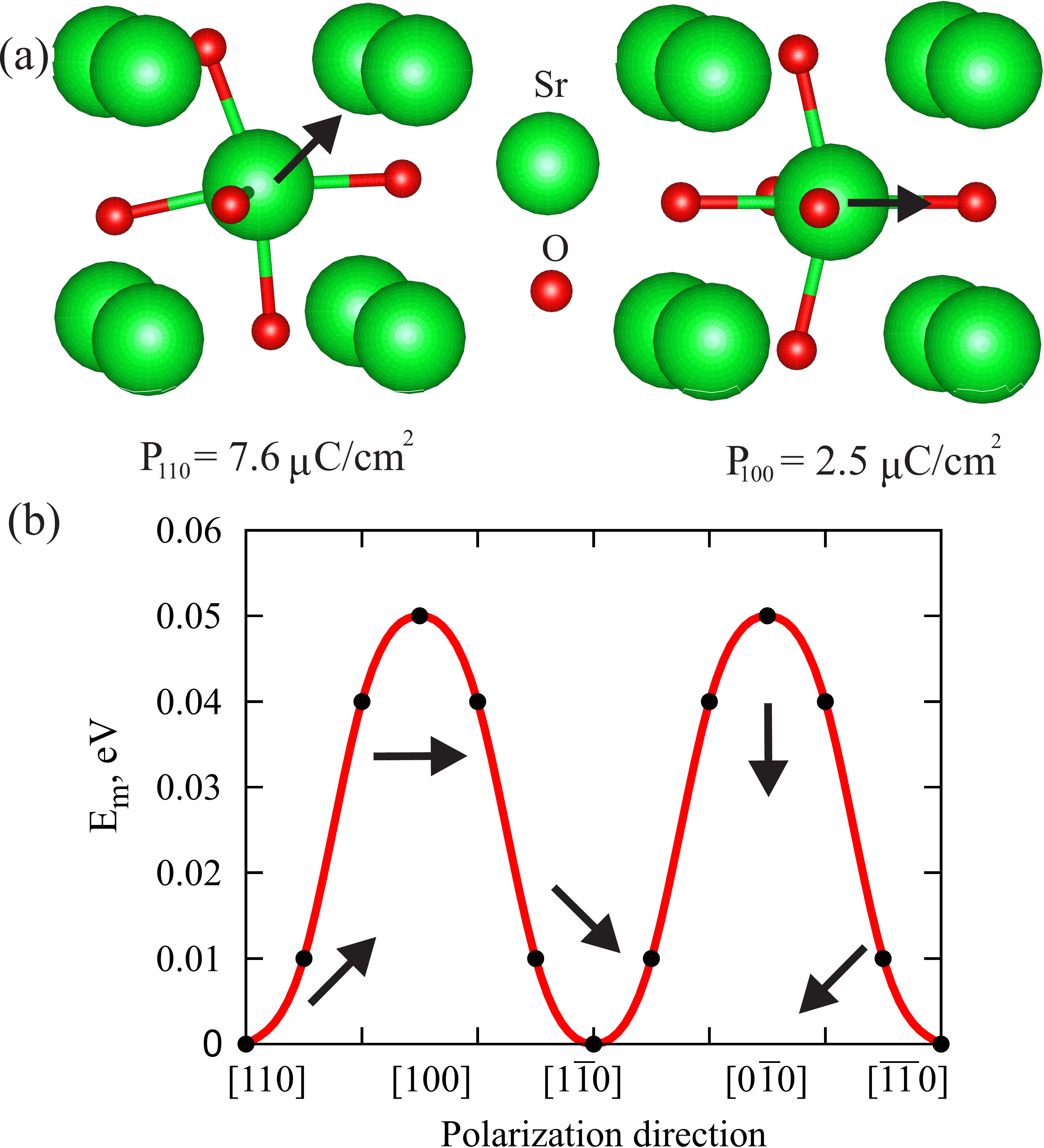}
  \caption{(a) Atomic structures of SrTiO$_3$ with the antisite Sr$_\mathrm{Ti}^{''}$ defect corresponding to two different polarization states with the defect shifted along [110] and [100] directions. (b) Migration energy profile between polarization states caused by the Sr$_\mathrm{Ti}^{''}$ defect. Polarization switching from [110] to [$\bar{1}$$\bar{1}$0] direction can be achieved via the polarization states with the [100] and [0$\bar{1}$0] directions.}\label{antiSr}
\end{figure}

Similarly to Ti$_\mathrm{Sr}^{\bullet \bullet}$, the formation of the antisite Sr$_\mathrm{Ti}^{''}$ defect in which a Sr ion substitutes one Ti ion is expected in Sr-rich SrTiO$_{3}$ (Figure \ref{antiSr}).\cite{liu2014composition} Our calculations reveal that the most energetically favorable configuration of Sr$_\mathrm{Ti}^{''}$ has an off-centering of 0.26 {\AA} along the [110] direction. The Sr$-$O distances become 2.22-2.26 {\AA} that are considerably shorter than those in pristine SrTiO$_3$ (2.76 {\AA}). This configuration can be regarded as an electric dipole composed of a strontium interstitial and a titanium vacancy. The calculated electric polarization $P^{110}$(Sr$_\mathrm{Ti}^{''}$) equals to 7.6 $\mu$C/cm$^2$ which is about twice smaller than in the Ti$_\mathrm{Sr}^{\bullet \bullet}$ case. The energy barrier calculated for polarization switching is only 0.05 eV rendering a low coercive voltage (Figure \ref{antiSr}). The contribution of the antisite Sr atom to the total dipole moment of the supercell is found to be about 10.6\% being comparable to the Ti$_\mathrm{Sr}^{\bullet \bullet}$ case. This spin-polarized structure of Sr$_\mathrm{Ti}^{''}$ induces magnetic moments on the nearest to Sr$_\mathrm{Ti}^{''}$  atoms and is more energetically favorable than the non-magnetic structure by about 0.17 eV exhibiting a much higher polarization switching barrier of $\sim$0.3 eV. We also estimate polarization $P^{100}$(Sr$_\mathrm{Ti}^{''}$) induced by the Sr$_\mathrm{Ti}^{''}$ displacement along the [100] direction which is the direction of film growth to be as low as 2.5 $\mu$C/cm$^2$ that may partially explain the absence of ferroelectricity in Sr-rich SrTiO$_3$ (001) thin films.\cite{yang2015room}

\begin{figure}[h!]
  \centering
  \includegraphics[width=0.4\textwidth]{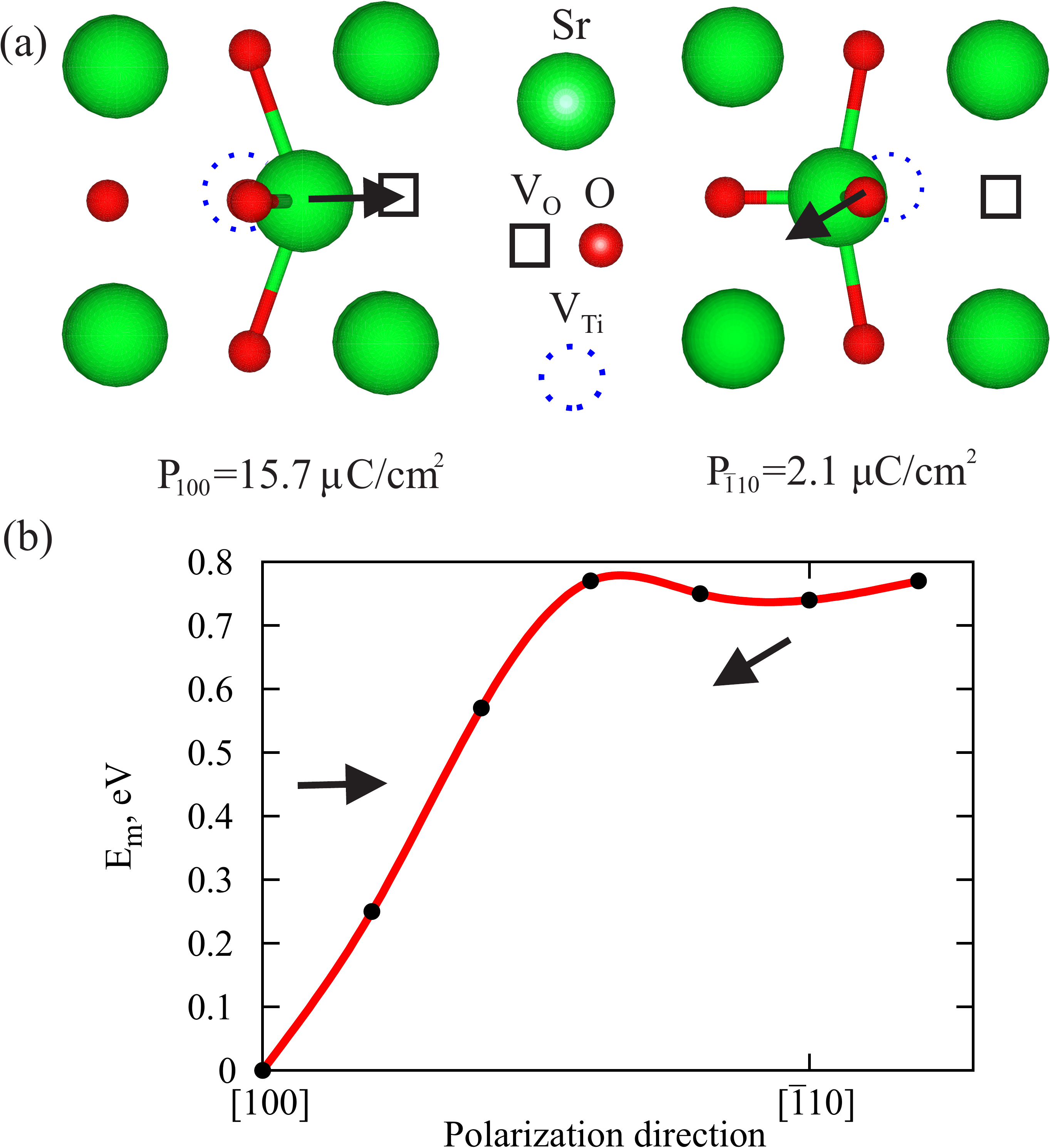}
  \caption{(a) Atomic structures of SrTiO$_3$ with the Sr$_\mathrm{Ti}^{''}$ defect and neutral V$_\mathrm{O}^{\times}$ corresponding to two different polarization states with the antisite defect shifted along the [100] and [$\bar{1}$10] directions. (b) Energy profile between two  polarization states caused by Sr$_\mathrm{Ti}^{''}$ and V$_\mathrm{O}^{\times}$. Polarization state for [$\bar{1}$10] direction has a very flat minimum suggesting that the state with Sr$_\mathrm{Ti}^{''}$ shifted along the [100] direction acts as a trap.}\label{antiSrO}
\end{figure}

The addition of oxygen vacancies is also found to have a significant impact on ferroelectric polarization induced by the Sr$_\mathrm{Ti}^{''}$ defect.
Recently, the formation of Sr$_\mathrm{Ti}^{''}$$-$V$_\mathrm{O}^{\bullet \bullet}$ defect complexes was observed experimentally during the electroforming and resistive switching of SrTiO$_{3}$.\cite{lenser2015formation} These complexes were previously calculated to have low formation enthalpies under Sr-rich conditions\cite{liu2014composition} and we estimate that the Sr$_\mathrm{Ti}^{''}$ defect has very large binding energies of -1.76 eV  and -1.85 eV with doubly charged V$_\mathrm{O}^{\bullet \bullet}$ and neutral V$_\mathrm{O}^{\times}$ vacancies, correspondingly.

Our calculations show that the positively charged oxygen vacancy causes a metallic state near the Fermi level and therefore no polarization can be provided for the Sr$_\mathrm{Ti}^{''}$$-$V$_\mathrm{O}^{\bullet \bullet}$ defect pair. On the other hand, neutral V$_\mathrm{O}^{\times}$ leads to semiconducting behavior and the most stable structure is characterized by a large off-centering (0.81 {\AA}) of the antisite defect along the [100] direction as shown in Figure \ref{antiSrO}. In this case the antisite Sr$_\mathrm{Ti}^{''}$ forms four short bonds of 2.23 {\AA} and one much longer bond of 2.72 {\AA} with the neighboring oxygen atoms.  The average polarization of the supercell is estimated as 15.7 $\mu$C/cm$^2$. The energy profile  of Sr$_\mathrm{Ti}^{''}$ diffusion associated with polarization switching in the presence of V$_\mathrm{O}^{\times}$  becomes  non-symmetrical with a very high switching barrier of 0.76 eV and a flat minimum for the $P_-$ state (Figure \ref{antiSrO}). This state induces a small polarization of 2.1 $\mu$C/cm$^2$ and should be unstable with respect to polarization switching. The switching via diffusion of oxygen vacancies is expected to have large barriers ($\sim$0.6-1.0 eV).\cite{lee2007oxygen}

In general, the results obtained for spontaneous polarization induced by the antisite Ti$_\mathrm{Sr}^{\bullet \bullet}$ and Sr$_\mathrm{Ti}^{''}$ defects are in qualitative agreement with experimental findings showing that although the excess of Sr can lead to ferroelectricity in polycrystalline SrTiO$_3$ at low temperatures, the observed polarization is considerably lower than for Ti-rich samples.\cite{guo2012ferroelectricity}

\subsection{Frenkel defects and small polarons}

\begin{figure}[h]
  \centering
  \includegraphics[width=0.4\textwidth]{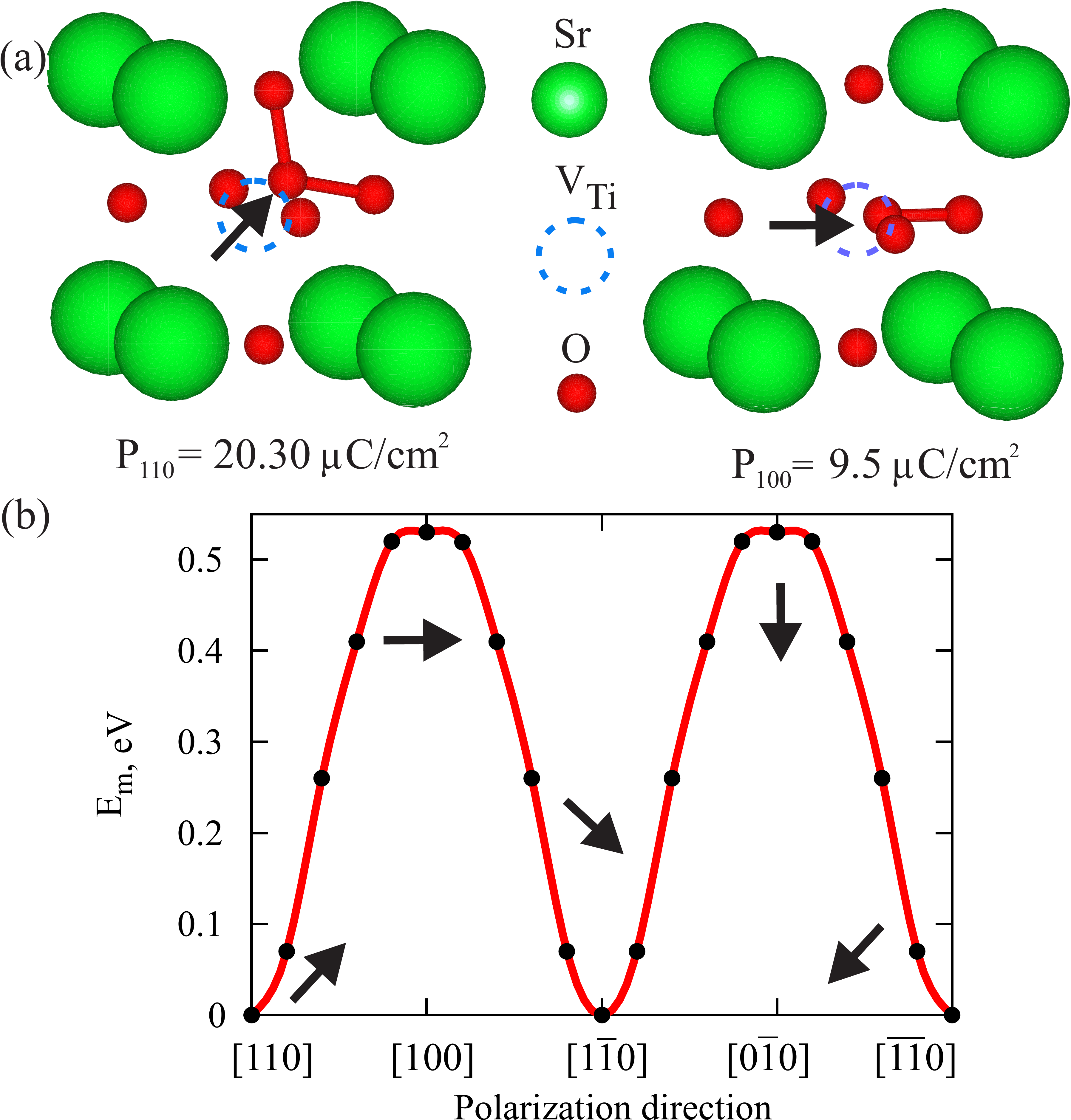}
  \caption{Atomic structures of SrTiO$_3$ with the Frenkel defect pair V$_\mathrm{Ti}^{''''}$$-$O$_\mathrm{i}^{\times}$ corresponding to two different polarization states with O$_\mathrm{i}^{\times}$ shifted along the [110]  and [100] directions. (b) Migration energy profile between polarization states caused by V$_\mathrm{Ti}^{''''}$$-$O$_\mathrm{i}^{\times}$.}
  \label{Tivac}
\end{figure}

The deficiency of cation atoms and excess of oxygen atoms leads to the formation of Frenkel defect pairs. In the case of titanium vacancy V$_\mathrm{Ti}^{''''}$ and oxygen interstitial O$_\mathrm{i}^{\times}$ pair we find that the most stable position for O$_\mathrm{i}^{\times}$ is to be shifted from the V$_\mathrm{Ti}^{''''}$ site along the [110] direction by 0.61 {\AA} as depicted in Figure \ref{Tivac}. The distance between O$_\mathrm{i}^{\times}$ and two adjacent lattice oxygen atoms is 1.35 {\AA}, while the corresponding angle between three oxygen atoms is about $110^{\circ}$. The electric dipole formed by this Frenkel pair causes a large average polarization $P^{110}$(V$_\mathrm{Ti}^{''''}$$-$O$_\mathrm{i}^{\times}$) of about 20.3 $\mu$C/cm$^2$, but with a high switching barrier of 0.54 eV.

\begin{figure}[h]
  \centering
  \includegraphics[width=0.4\textwidth]{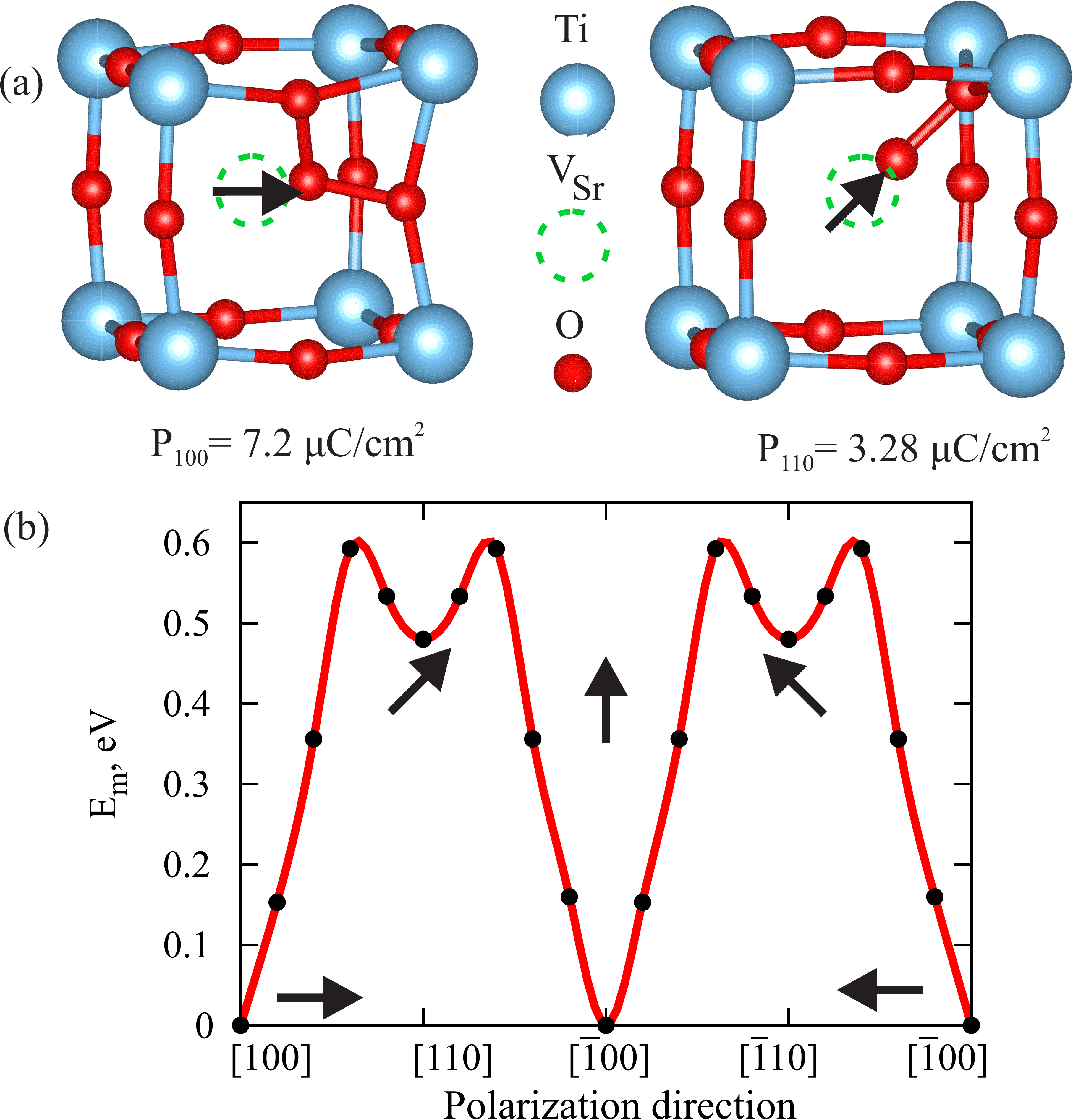}
  \caption{ Atomic structures of SrTiO$_3$ with the Frenkel defect pair V$_\mathrm{Sr}^{''}$$-$O$_\mathrm{i}^{\times}$ corresponding to two different polarization states with O$_\mathrm{i}^{\times}$ shifted along the [100] and [110] directions. (b) Migration energy profile between polarization states caused V$_\mathrm{Sr}^{''}$$-$O$_\mathrm{i}^{\times}$.}
   \label{Srvca}
\end{figure}

Calculations of the other Frenkel defect pair composed of a Sr vacancy and an oxygen interstitial reveal that it is energetically preferable for O$_\mathrm{i}^{\times}$  to be shifted along the [100] direction with the 1.24 {\AA} off-centering from the initial Sr position (Figure \ref{Srvca}). However, such a significant off-centering does not induce a large local dipole moment because of the very small Born charge of 0.15 on the O interstitial (see Table \ref{tab1}). The overall polarization of the supercell in this case is computed to be around 7.2 $\mu$C/cm$^2$  with the high diffusion barrier for polarization switching of 0.61 eV.

It was previously shown that excess electrons in the bulk SrTiO$_{3}$ do not become localized in the form of small polarons on Ti atoms, but can be stabilized in the presence of oxygen vacancies.\cite{polaron} It turned out that in $n$-type SrTiO$_{3}$ the most stable configuration corresponds to the case when each oxygen vacancy traps one small polaron remaining in a +1 charge state and providing one electron to the conduction band. We find that the dipole moment produced by such a defect pair causes a moderately large polarization of 5.0~$\mu$C/cm$^2$.

\subsection{The impact of defect concentration and the SrTiO$_3$/SrRuO$_3$ interface}

\begin{figure}[!h]
  \centering
  \includegraphics[width=0.4\textwidth]{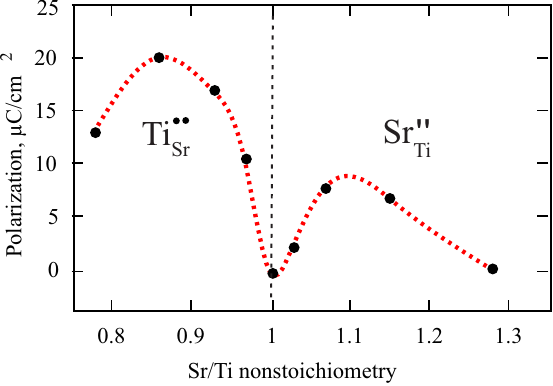}
  \caption{Average spontaneous polarization as a function of defect concentration. Sr-rich condition Sr/Ti $>$ 1 corresponds to the larger concentration of Sr$_\mathrm{Ti}^{''}$ and Sr/Ti $<$ 1 corresponds to the larger concentration of Ti$_\mathrm{Sr}^{\bullet \bullet}$.}
   \label{concentration}
\end{figure}

In this section we aim to examine how the defect concentration and the presence of the interface with SrRuO$_{3}$ can impact polarization properties of SrTiO$_{3}$.
To simulate different concentrations  of  the antisite Ti$_\mathrm{Sr}^{\bullet \bullet}$ and Sr$_\mathrm{Ti}^{''}$ defects we consider one defect in 2$\times$2$\times$2, 3$\times$3$\times$3 and 4$\times$4$\times$4 supercells corresponding to Sr/Ti  ratio of 0.78, 0.93, 0.97, 1.03, 1.07 and 1.28,  respectively.  In addition, we examine two Ti$_\mathrm{Sr}^{\bullet \bullet}$ (or Sr$_\mathrm{Ti}^{''}$) defects in a $3\times 3\times 3$ supercell with the largest defect separation attainable in this cell which corresponds to  Sr/Ti ratio of 0.86 and 1.16.  As it seen from Figure \ref{concentration}, an increase of the Ti$_\mathrm{Sr}^{\bullet \bullet}$ defect concentration causes noticeably enhanced polarization, but as the defect concentration increases polarization gets diminished partly due to a much smaller displacement of Ti$_\mathrm{Sr}^{\bullet \bullet}$ being 0.45 {\AA} for Sr/Ti = 0.78 as compared to 0.78 {\AA} for Sr/Ti = 0.93. A similar trend is observed for the Sr$_\mathrm{Ti}^{''}$ defect and we also find that the high concentration of antisite Sr$_\mathrm{Ti}$ (Sr/Ti = 1.29) leads to metallic electronic structure. Importantly, for a Sr/Ti ratio of 1.16 the system with two neighboring  Sr$_\mathrm{Ti}^{''}$ defects become more stable if the defects are displaced along the different directions ([110] and $[1\bar{1}0]$) giving rise to a decrease of the total polarization, the effect that is not observed for Ti$_\mathrm{Sr}^{\bullet \bullet}$. Overall, we predict the same trend for spontaneous polarization as a function of Sr/Ti nonstoichiometry as previously measured for Ti- and Sr-rich SrTiO$_{3}$ samples,\cite{guo2012ferroelectricity} with the antisite Ti$_\mathrm{Sr}^{\bullet \bullet}$ defect causing a more pronounced polarization than Sr$_\mathrm{Ti}^{''}$ for the same defect concentration.

\begin{table}[h]
\caption{Quantities calculated for a $3\times 3\times 3$ SrTiO$_3$  supercell with different defects:  defect off-centering \emph{d} along the corresponding directions, Born charge associated with the off-centered cation, average spontaneous polarization \emph{P}, activation barrier for polarization switching $E_a$. Calculated Born charges for pristine SrTiO$_3$ are 2.56, 6.57, -5.23 and -1.93 for Sr, Ti, O$_{\parallel}$ and O$_{\perp}$, correspondingly.}
\begin{tabular}{ccccc}
  \hline
  \hline
  % after \\: \hline or \cline{col1-col2} \cline{col3-col4} ...
     Defect & \emph{d}(\text{\AA}) & Born charge & \emph{P}($\mu C/cm^2$) &   \emph{E}$_m$(eV)  \\
  \hline
  Ti$_\mathrm{Sr}^{\bullet \bullet}$ & 0.78 [001] & 1.72 & 16.8 & 0.13 \\
  Ti$_\mathrm{Sr}^{\bullet \bullet}$-V$_O^{\times}$ & 0.82 [011] & -- & -- & -- \\
  Ti$_\mathrm{Sr}^{\bullet \bullet}$-V$_O^{\bullet \bullet}$ & 0.79 [011] & 2.48 & 22.6 & 0.23 \\
  Sr$_\mathrm{Ti}^{''}$ & 0.26 [011] & 3.11 & 7.6 & 0.05  \\
  Sr$_\mathrm{Ti}^{''}$-V$_O^{\times}$ & 0.81 [001] & 3.59 & 15.7 & 0.76  \\
  Sr$_\mathrm{Ti}^{''}$-V$_O^{\bullet \bullet}$ & 0.81 [001] & -- & -- & --  \\
  V$_\mathrm{Ti}^{''''}$-O$_i^{\times}$ &  0.61 [110] & 2.2 & 20.3 & 0.54 \\
  V$_\mathrm{Sr}^{''}$-O$_i^{\times}$ & 1.24 [001] & 0.15 & 7.2 & 0.61 \\
  Ti$_\mathrm{Ti}^{\bullet}$-V$_O^{\times}$ & 0.08 [001] & 5.1 & 5.0 & -- \\
  \hline
  \hline
\end{tabular}
\label{tab1}
\end{table}

To obtain some insight into the impact of thin-film interface on polarization properties, we focus on the antisite Ti$_\mathrm{Sr}^{\bullet \bullet}$ defect that exhibits the most pronounced and easily switchable polarization in the bulk phase. It was previously demonstrated that the creation of this defect in the SrTiO$_3$/SrRuO$_3$ thin films is more probable than in the bulk SrTiO$_3$ due to its lower formation energy.\cite{lee2015emergence} Since no polarization was experimentally detected in SrRuO$_3$ region of the heterostructure,\cite{lee2015emergence} we assume that all the dipole moments are induced by the four SrTiO$_3$ layers.

In order to directly compare spontaneous polarization of the SrTiO$_3$/SrRuO$_3$ interfacial structure with the case of bulk SrTiO$_3$, we also estimate polarization for a  $3\times 3\times 4$ supercell of the bulk SrTiO$_3$ that corresponds to the same number of SrTiO$_3$ layers as in the heterostructure. Our calculations predict that the presence of the interface with metallic SrRuO$_3$ leads to a reduction of the average polarization from 13.3 $\mu$C/cm$^2$ for the bulk down to 8.1 $\mu$C/cm$^2$ for the thin film. Based on the obtained results and the fact that the formation energy of Ti$_\mathrm{Sr}^{\bullet \bullet}$ becomes significantly reduced in thin films,\cite{lee2015emergence} we conclude that the enhancement of polarization in thin films should occur due to the high concentration of defects rather than the influence of the SrTiO$_3$/SrRuO$_3$ interface.

\section{Conclusions}

In summary, we have explored the impact of a range of native point defects on ferroelectric polarization and the mechanisms of polarization reversal in bulk and thin films of SrTiO$_{3}$ by employing DFT calculations in combination with the Berry phase approach. We have shown that the antisite Ti$_\mathrm{Sr}^{\bullet \bullet}$ defect should result in the pronounced spontaneous polarization, however, the presence of oxygen vacancies may substantially reduce the polarization, make polarization switching barriers much higher and even cause non-insulating behavior. The presence of antisite Sr$_\mathrm{Ti}^{''}$ induces smaller polarization with lower barriers of polarization switching than those for Ti$_\mathrm{Sr}^{\bullet \bullet}$, in quantitative agreement with previously measured polarization for Sr- and Ti-rich SrTiO$_{3}$ samples. We have also found that the increase in spontaneous polarization in SrTiO$_3$/SrRuO$_3$ thin films can be achieved by tailoring the degree of Sr/Ti nonstoichiometry and is not due to the presence of SrTiO$_3$/SrRuO$_3$ interfaces. Some other intrinsic point defects such as Frenkel defect pairs and electron small polarons have been also found to give sizable contributions to spontaneous polarization of SrTiO$_{3}$.

\begin{acknowledgments}
We would like to thank Alexey Gruverman for fruitful discussions and comments on this study.
The Holland Computing Center at the University of Nebraska-Lincoln is acknowledged for computational support. This work was supported by the National Science Foundation (NSF) through the Nebraska Materials Research Science and Engineering Center (MRSEC) (grant No. DMR-1420645). V.A. gratefully acknowledges support from the startup package.
\end{acknowledgments}

\bibliography{main}

\end{document}